# High Discrimination Ratio, Broadband Circularly Polarized Light Photodetector Using Dielectric Achiral Nanostructures

Guanyu Zhang[1], Xiaying Lyu[1], Yulu Qin[1], Yaolong Li[1], Zipu Fan[2], Xianghan Meng[1], Yuqing Cheng[3], Zini Cao[1], Yixuan Xu[1], Dong Sun[2], Yunan Gao[1, 4, 5], Qihuang Gong[1, 4, 5], and Guowei Lu[1, 4, 5*]

*[1] State Key Laboratory for Mesoscopic Physics, Collaborative Innovation Center of Quantum Matter, Frontiers Science Center for Nano-Optoelectronics, School of Physics, Peking University, Beijing 100871, China*

*[2] International Center for Quantum Materials, School of Physics, Peking University, Beijing 100871, China*

*[3] School of Mathematics and Physics, University of Science and Technology Beijing, Beijing 100083, China*

*[4] Collaborative Innovation Center of Extreme Optics, Shanxi University, Taiyuan, Shanxi 030006, China*

*[5] Peking University Yangtze Delta Institute of Optoelectronics, Nantong, Jiangsu 226010, China*

*\* Corresponding author: Guowei.lu@pku.edu.cn (G. Lu)*

## Abstract

The on-chip measurement of polarization states plays an increasingly crucial role in modern sensing and imaging applications. While high-performance monolithic linearly polarized photodetectors have been extensively studied, integrated circularly polarized light (CPL) photodetectors are still hindered by inadequate discrimination capability. In this study, we employ achiral all-dielectric nanostructures to develop a broadband CPL photodetector with an impressive discrimination ratio of ~107 at the wavelength of 405 nm, significantly surpassing its counterparts by two orders of magnitude. Our device shows outstanding CPL discrimination capability across the visible band without requiring intensity calibration. Its function mechanism is based on the CPL-dependent near-field modes within achiral structures: under left or right CPL illumination, distinct near-field modes are excited, resulting in asymmetric irradiation of the two electrodes and generating a photovoltage with directions determined by the chirality of the incident light field. The proposed design strategy facilitates the realization of ultra-compact CPL detection across diverse materials, structures, and spectral ranges, presenting a novel avenue for achieving high-performance monolithic CPL detection.

## Introduction

Circularly polarized light (CPL), also known as spin light, plays an important role in various contemporary applications such as chiral molecule distinguishing[1], remote sensing[2], quantum optics[3], and spintronics[4]. Traditional methods for CPL detection require bulky optical components to convert them into linearly polarized light, which hinders device miniaturization and fails to meet the evolving demands of on-chip polarimeter applications. Although replacing bulk optics with ultrathin metasurfaces can reduce the size of the detection system, maintaining a certain propagation distance between the metasurface and detector array remains essential.[5] Additionally, complex fabrication processes for composite metasurfaces pose significant challenges for cost-effective production of monolithic CPL photodetectors. Consequently, an integrated, high-performance CPL detection scheme is urgently desired.

Three main approaches are known to achieve filterless CPL photodetectors: chiral materials, novel photoelectric mechanisms, and artificial structures. Optoelectronic materials exhibiting circular dichroism have been widely proposed for high-responsivity, low-cost light ellipticity detection[6–10]. However, the ability of these devices to distinguish CPL is typically compromised due to the inherent weak chiroptic response in natural molecules. Furthermore, the material's circular dichroism peak limits its operational wavelength range, resulting in a narrow bandwidth constraint. Novel photoresponse mechanisms such as the circular photogalvanic effect[11–13], photon drag effect[14], spin-locked photocurrent in topological insulators[15,16], and spin-galvanic effect[17] can generate spin-dependent photoelectric responses. Nevertheless, these mechanisms currently exhibit extremely low responsivities and are often overshadowed by other competing photoresponse mechanisms, thereby hindering their practical applications.

Artificial structures have been proven to exhibit strong chirality several orders surpassing that of natural materials.[18–20] The integration of plasmonic chiral structures with optoelectronic materials such as graphene[21], $MoS_2$[22], and silicon[23], facilitates the efficient conversion of light field chirality into electrical readouts. This approach has led to the development of compact CPL photodetectors with high responsivity, superior discrimination ratio, and extensive design flexibility in near and mid-infrared band[24–26]. Despite these advancements, the expansion of metal metasurface-based CPL photodetectors to the visible spectrum has not yet been achieved, potentially due to their large intrinsic loss and fabrication difficulties. Dielectric metasurfaces directly etched on optoelectronic materials present a promising solution[27]. Nanostructures with broken mirror symmetry exhibit different absorption under opposite CPL, resulting in distinct photoresponse intensities. However, achieving such symmetry broken in the z-direction with uniform dielectric substrates remains challenging, leading to inadequate intrinsic chirality. This weak inherent chirality compromises the discrimination ability of devices based on dielectric nanostructures[28].

Generally, discrimination ratio (DR) is employed to quantify a device's capability in distinguishing between left-circularly polarized (LCP) and right-circularly polarized (RCP) light. It is defined by the equation: DR= $2|(R_{LCP} - R_{RCP})/(R_{LCP} + R_{RCP})|$), which measures the absolute difference in photoresponses to LCP and RCP light relative to their average value. Currently, CPL photodetectors operating within the visible spectrum typically exhibit a DR <1,[27] significantly below the threshold for practical applications (Fig. 1d) . Moreover, the photoresponse magnitudes in these absorption-based devices are determined by both the light intensity and its polarization state, making them sensitive to interference from light intensity fluctuations and necessitating calibration before use. These factors highlight an urgent demand for high-performance CPL detectors that operate within the visible spectrum.

In this work, by introducing achiral dielectric nanostructures, we demonstrate a high discrimination ratio CPL photodetector capable of operating across the entire visible spectrum, which uniquely determines the chirality of the light field through the polarity of the photovoltage. We have extended the phenomenon of hidden chirality[29,30], previously explored in plasmonic structures, to dielectric metasurfaces, thereby enabling a CPL-sensitive near-field mode. This mode is converted to chirality-dependent photovoltage through photovoltaic (PV) or photothermoelectric (PTE) effects. The near-field modes of achiral dielectric nanostructures are numerically analyzed and experimentally validated using photoemission electron microscopy (PEEM). A high-discrimination ratio CPL photodetector operating across the visible spectrum is developed using tellurium (Te) nanosheets, achieving an unprecedented maximum DR of 107 and responsivity of 0.368V/W. This device exhibits exceptional DR across the visible spectrum, with the direction of photoresponse providing a direct measurement of the light field's chirality, free from the light intensity calibration. These breakthroughs position our device as a compelling candidate for ultracompact, high-performance CPL detectors.

## Results

**Methodology**

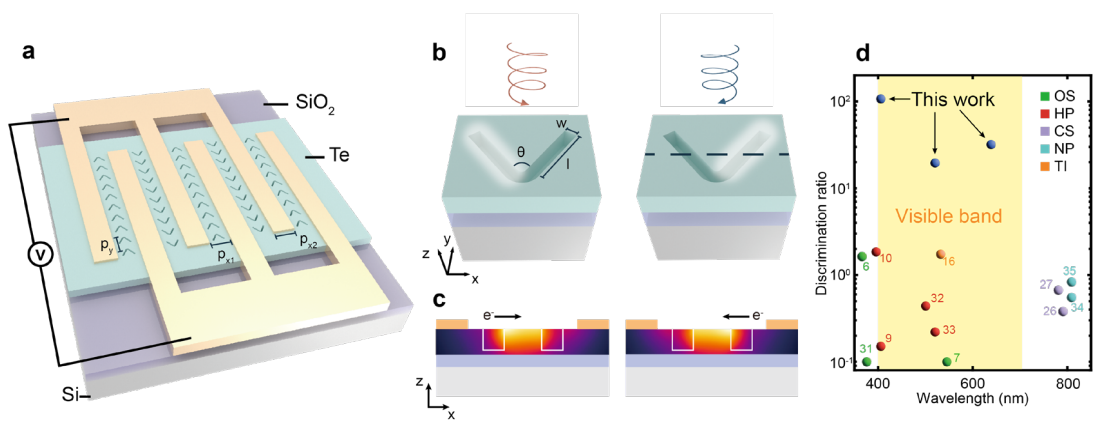

**Fig. 1 Structure and working principle of the high discrimination ratio CPL detector. a.** Structure of an array device with a pair of interdigital electrodes and V-grooves situated between them. The basic unit consisted of two electrodes is a unit device. The electrode width and spacing are $p_{x1}=p_{x2}$=800 nm, and the structure spacing is $p_y$=480 nm. **b.** Working principle of the device. Under LCP (RCP) light illumination, near-field modes will be localized at the right (left) arms of V-grooves. *l*, *w*, and *θ* represent arm length, width, and angle, respectively. **c.** The x-z view of a unit device and the thermal simulation at the cross-section line in the above figures. **d.** Comparison of the achiral-structure-based device to other visible band CPL photodetectors. Here, organic semiconductors(OS), hybrid perovskites(HP), chiral structures(CS), chiral nanoparticles(NP), and topological insulators(TI) imply the detection mechanisms of these devices[6–10,16,26–28,31–35].

Fig. 1a presents the structure of our CPL photodetector. The basic unit of the device comprises a pair of electrodes positioned on Te nanosheets, with a series of etched V-shaped grooves symmetrically situated between them. Array devices are dual-port devices composed of multiple basic units, which will be discussed later. Unless otherwise stated, all measurements in this study were conducted under zero bias.

The device operates through chirality-resolved near-field modes in achiral structures, leveraging a phenomenon known as hidden chirality. Contrary to the conventional belief that chiral geometries are required for CPL discrimination, recent advancements in plasmonic metasurfaces have demonstrated that achiral structures can exhibit different near-field modes under LCP and RCP light, despite their identical far-field responses[24,36]. This concept has been extended to dielectric nanostructures, as illustrated in Fig. 1b and 2b. Specifically, in carefully designed V-grooves, exposure to LCP or RCP light results in light localization at the right or left arms, respectively, leading to photoresponses of equal magnitude but opposite directions. This chiral-dependent mode induces differential temperature increases on either side of the V-grooves, depicted in Fig. 1c. By integrating electrodes on both sides of the V-groove and employing the PTE effect, this temperature difference generates photovoltage in opposing directions. The broadband nature of these near-field modes endows the device with broadband response capabilities. Such mode arises from the interference of x-polarized (LP x) and y-polarized (LP y) light excitation modes, depicted in Fig. 2b. When LP x and LP y light excite the structure, the resulting near-field modes distribute symmetrically, resulting in zero net photovoltage. Under CPL light, these two modes are mixed with different phases, enabling chirality-dependent near-field patterns.

The CPL detection scheme discussed is not limited to specific materials, optoelectronic mechanisms, or geometry structures, thus offering great versatility. In this study, Te nanosheets were employed as the photothermoelectric material due to their straightforward synthesis process, high refractive index, and superior thermoelectric properties[37,38]. Otherwise, any material with a sufficient refractive index to support Mie resonance modes and capable of effectively converting light into a photoelectric response can facilitate CPL detection through achiral structures. Apart from PTE, any mechanism capable of inducing a photoelectric response at the electrode-material interface is applicable, such as the photovoltaic effect widely existing in the semiconductor-metal and semiconductor-semiconductor interface. For instance, achiral silicon structures under CPL excitation can exhibit chirality-sensitive modes that asymmetrically illuminate the electrodes, causing an uneven distribution of photogenerated carriers on two sides. These carriers are then separated by the built-in field of the Schottky junction, resulting in a photoresponse directed by the light chirality. Furthermore, the utilization of V-shaped geometric structures is not the sole alternative. Any geometry that maintains a singular symmetry plane, such as T-shaped grooves (see Supplementary Figure S1), could fulfill a similar function. The flexibility of this achiral-geometry-based CPL detection method opens up a broad spectrum of potential applications.

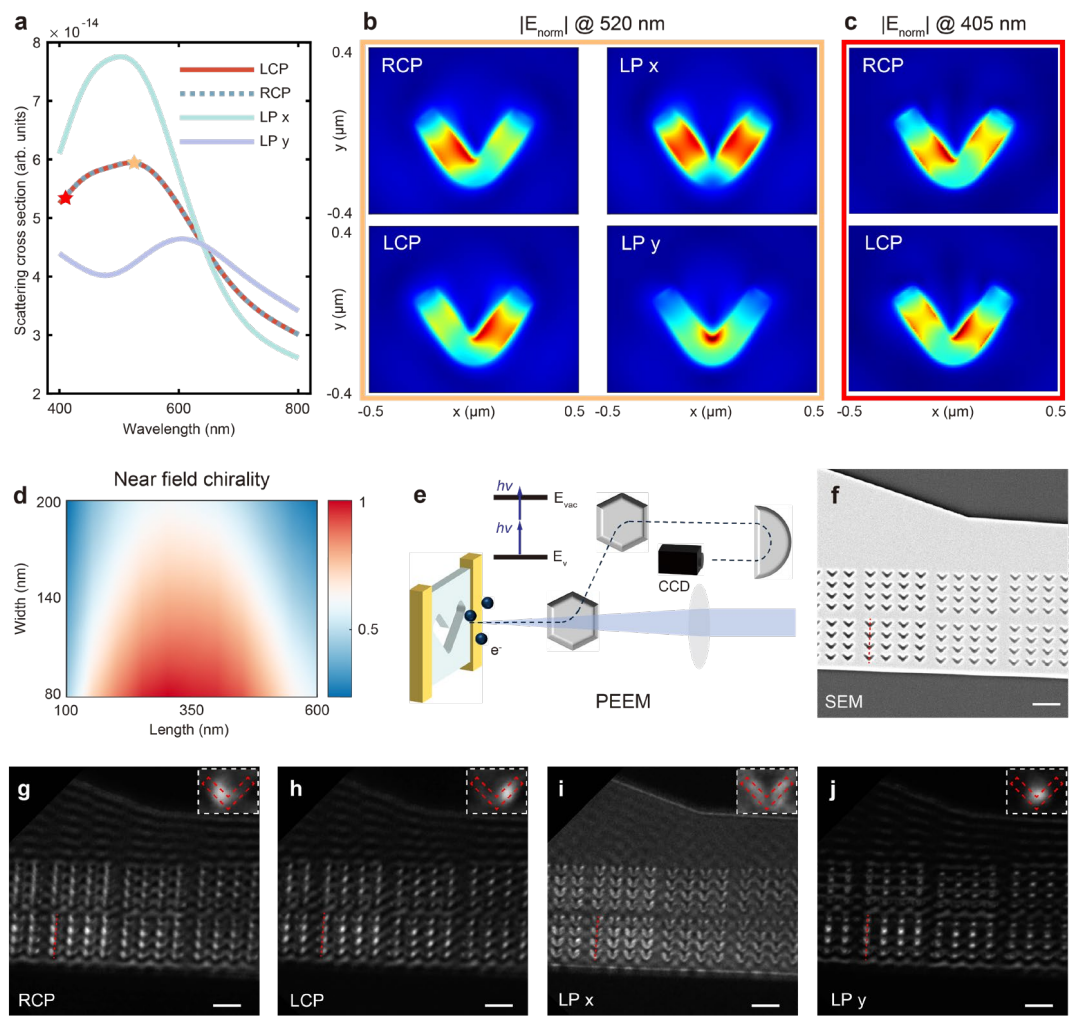

**Fig. 2 Near field modes of V-grooves. a.** Scattering cross section of optimized V-grooves under LCP, RCP, LP x, and LP y light. **b.** Normalized electric field intensity $|E_{norm}|$ of RCP, LCP, LP x, LP

y light excitation modes on V-shaped grooves at the wavelength of 520 nm. Normalization is performed at the maximum intensity under 520 nm. **c.** Normalized electric field intensity of the modes excited by RCP and LCP light at 405 nm. Normalization is performed at the maximum intensity under 405 nm. **d.** Geometric parameter optimization of V-groove arm length and width. Normalized near-field chirality is set as the evaluation metric. **e.** Schematic diagram of the working principle of PEEM. **f.** SEM image of the V-groove arrays corresponding to PEEM results. **g-j.** The near-field mode of V-groove arrays observed with PEEM under a 410 nm laser. Figures from left to right represent modes excited by RCP, LCP, LP x, LP y light. The red vertical dashed line is located in the center of a row of V-grooves to facilitate mode observation. The enlarged view of a single V-groove is in the upper right corner figure, whose edge is indicated by the red wireframe. All scale bars in Fig. 2f-j are 1 μm.

**Design of the achiral nanostructure**

The geometric parameters of the nanostructure were optimized through numerical simulations. Near field chirality, defined as the difference between the mean light intensities on the structure's left and right sides under CPL excitation, serves as a metric to assess the discrimination capacity of geometric structures. The optimized geometric parameters of the V-groove for 520 nm wavelength are arm length $l$=300 nm, arm width $w$=100 nm, and the included angle $\theta$ =70°, as shown in Fig. 2d and Supplementary Figure S2. Additionally, we investigated the influence of Te nanosheet thickness and groove depth on near-field chirality. According to the calculations, increasing groove depth enhances chiral resolution, necessitating thicker nanosheets.

We further conducted a mode analysis of the V-shaped groove. Fig. 2a illustrates the resonance modes of the optimized V-shaped groove when subjected to various polarized lights. As expected, the achiral nanostructure shows an identical scattering cross-section under the LCP and RCP light. Typically, the wavelength shift observed in resonance peaks under LP x and LP y light indicates near-field chirality, which is mutually confirmed with the aforementioned parameter optimization results. The near-field modes of the V-shaped groove under various wavelengths and polarizations are shown in Supplementary Figure S3, 4. Under the CPL excitation, the light field focuses on one arm of the V-shaped groove, showing highly asymmetric, whereas linearly polarized light yields symmetrical near-field modes. Under LP x light, the light field distributes evenly on both arms of the V-shape. With LP y light excitation, the field localizes at the V-shaped groove's apex. Such mode feature spans the entire visible and near-infrared band, endowing the device with a theoretical broadband response capability (Fig. 2c and Supplementary Figure S3). In structures with weak chiral resolution ($l$=500 nm, $w$=100 nm), negligible wavelength discrepancy exists between the resonance peaks for LP x and LP y light, suggesting an absence of chirality-dependent mode interference (Supplementary Figure S5).

To verify the numerical calculations, PEEM was employed for near-field measurement. The operational principle of PEEM is depicted in Fig. 2e. The light interacts with the sample, generating free electrons via the two-photon photoelectric effect. These electrons are then imaged onto a CCD after passing through a series of electromagnetic lenses, enabling high spatial resolution detection of near-field modes. The near-field modes corresponding to various polarizations are illustrated in Fig. 2g-j, aligned closely with simulations. Fig. 2f presents a scanning electron microscope (SEM) image of the corresponding area, showcasing V-shaped groove arrays with depths incrementing from right to left. As expected, deeper grooves correlate with heightened asymmetry in the near-field mode under CPL excitation. Notably, a 410 nm laser beam was employed in the PEEM measurements to fulfill the necessary condition for electrons to surpass the vacuum energy level.

**CPL-sensitive photovoltage in unit devices**

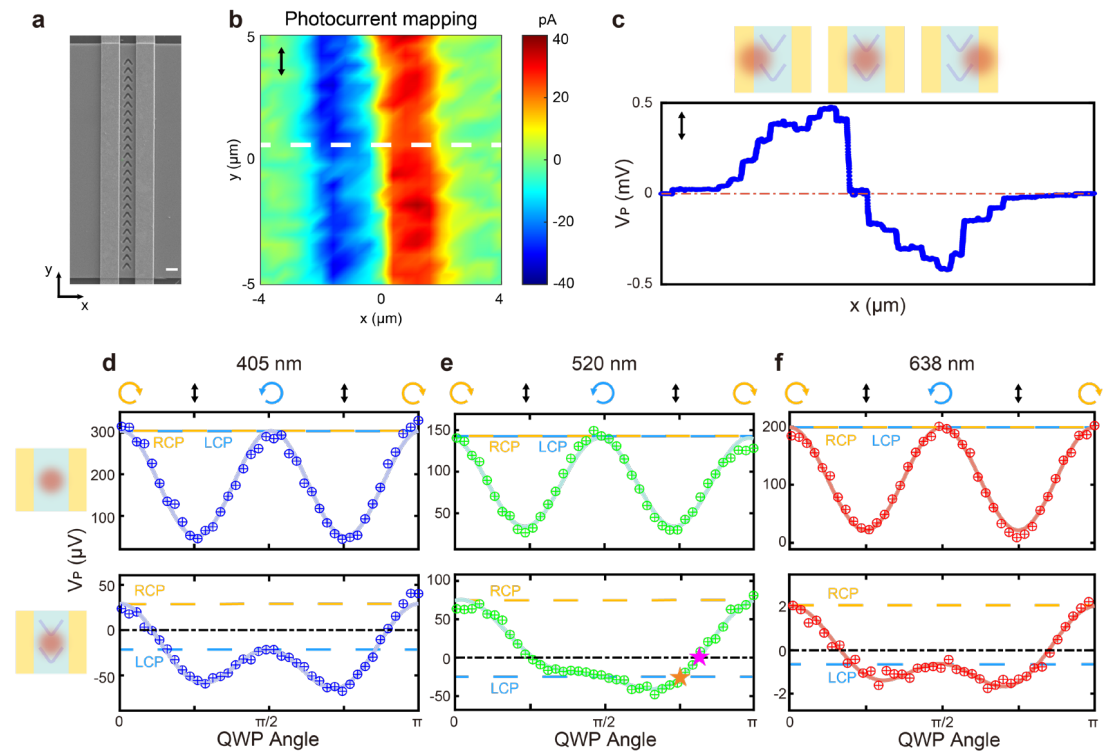

**Fig. 3 Optoelectronic characterization of unit devices. a.** SEM image of a unit device. The scale bar is 1 μm. **b.** Photocurrent mapping of a unit device under LP y light. **c.** Photovoltage line scan at the dashed line in Fig. 3b under LP y light. The upper diagram shows the relative position of the light spot and the device, corresponding to the change in photovoltage. **d-f.** Variation curves of photovoltage with QWP rotation angle at wavelengths of 405 nm, 520 nm, and 638 nm. The figures'

upper sections display results from devices without V-grooves, whereas the lower sections illustrate outcomes post-etching. The orange and magenta stars in Fig. 3e represent the transition point of the light chirality and the photovoltage polarity. Yellow and blue dashed lines represent photovoltage under RCP and LCP light, respectively.

An SEM image of our device is shown in Fig. 3a. Te nanosheets were synthesized by hydrothermal method[39] and dropped to $SiO_2$-Si substrate. The electrodes were fabricated through electron beam lithography, followed by focused ion beam (FIB) etching to create V-shaped grooves onto the Te nanosheets. The electrodes were oriented perpendicular to the Te nanosheets' long axis, corresponding to the direction of Te atomic chains, as confirmed by angle-resolved Raman spectroscopy (Supplementary Figure S6). Laser beams with wavelengths of 405 nm, 520 nm, and 638 nm were directed through a polarizer and an achromatic quarter-wave plate (QWP), then concentrated onto the devices with focused spot diameters around 1 μm. Devices were mounted on a piezo displacement stage. Scanning this stage enables photocurrent mapping, as shown in Fig. 3b. The photoresponse is believed to originate from the PTE effect, which is indicated by the linear power-dependence curve. Electrical tests confirm the absence of Schottky junctions between the electrodes and Te nanosheets (Supplementary Figure S7), ruling out the possibility of a built-in electric field induced photocurrent, consistent with previous literature[37]. When the light spot position moves perpendicular to the electrodes, it induces photoresponses in opposite directions on each side. With the light spot positioned between the electrodes—where the y-polarization response turns to zero (see Fig. 3c)—the device obtains an equilibrium state and exhibits photovoltage directed by the light field's chirality. During this phase, both linearly polarized and unpolarized light induce counteracting photoresponses on the electrodes, allowing the CPL-sensitive photovoltage to dominate. The light polarization state can be continuously modulated by rotating the QWP, facilitating polarization-dependent photovoltage measurements. The photovoltage curves varying with the QWP rotation angle φ in the equilibrium state are shown in Fig. 3d-f. The top panels show the outcomes before FIB etching, while the bottom panels show the results after etching V-grooves.

For the three testing wavelengths, the devices demonstrate photovoltage in contrasting directions under LCP and RCP light, differing from unetched devices' responses. A high-discrimination, broadband, calibration-free CPL photodetector has been realized. Notably, despite the intrinsic chirality of one-dimensional Te atomic chains, which are expected to cause varying absorption for different CPL[40]. The interaction between chiral light and these atomic structures is extremely weak in the visible band. Our experiments showed that unpatterned devices failed to demonstrate any appreciable circular polarization response. The device has a CPL responsivity of 0.37 V/W at 520 nm and a DR of 4.1, significantly surpassing the typical DR of reported visible CPL detectors.

Although the CPL-sensitive component, $V_C$, dominates the photoresponse at the equilibrium position, the overall photovoltage includes a polarization-insensitive term, $V_0$, and a linear polarization-sensitive term, $V_L$. The whole photoresponse is expressed as $V_{ph} = V_0 + V_L \cos(2\varphi+\psi_1) + V_C \cos(4\varphi+\psi_2)$, which accurately describes the experimental observations. $\psi_1$ and $\psi_2$ are phase constants. $V_0$ and $V_L$ cause a discrepancy in the photovoltage amplitudes induced by LCP and RCP light, yielding a limited DR. Additionally, the polarization-dependent curves reveal that photovoltage transition points do not fully align with chirality transition points, indicating that the sign of the photovoltage cannot completely reflect the chirality of the elliptically polarized light, as shown in Fig. 3e. To further suppress the $V_0$ and $V_L$ terms, their origins were carefully analyzed.

Notably, the polarization-insensitive component of photovoltage emerges from the incomplete elimination of the asymmetrical response generated by the incident light at the electrodes. Theoretically, a centrally positioned light spot should result in equal light intensity at both electrodes for a device without V-grooves. This leads to photovoltage of identical magnitude but opposite directions, achieving an equilibrium state with zero net photovoltage. However, in practice, disparities in electrode-nanosheet interfaces, fabrication imperfections, and alignment inaccuracies between the spot and the device prevent the total elimination of the net photovoltage (Supplementary Figure S8). Additionally, alignment inaccuracies during V-groove fabrication may offset the structure from the device's central axis, causing LCP and RCP photoexcitation to impact the electrodes unevenly. These factors jointly contribute to the formation of the polarization-insensitive component of photoresponse.

The linear polarization sensitive term $V_L$ has more complex origins, mainly derives from the modulation of baseline voltage $V_0$ induced by the differential absorption of variously polarized lights. Two competing mechanisms play a role in this absorption modulation: scattering at the gold electrode edges and the anisotropic absorption of Te material. Te nanosheets, composed of one-dimensional atomic chains bonded by van der Waals forces, exhibit strong anisotropy. Typically, light polarized perpendicular to the atomic chains, that is, LP y light, has a higher absorption coefficient[41]. On the contrary, the scattering effect at the electrode edges favors LP x light, facilitating more efficient coupling of light polarized perpendicular to the electrodes into the device. Generally, the scattering effect of the electrodes is the dominant factor, offsetting the absorption differences in Te, resulting in smaller photovoltages under LP y light excitation, as shown in Figure 3d-f and Supplementary Figure S9. Moreover, V-grooves demonstrate uneven near-field distributions when exposed to obliquely polarized light, which influences the $V_L$ term. However, it is not the primary contributor (Supplementary Figure S10).

**Performance optimization with array devices**

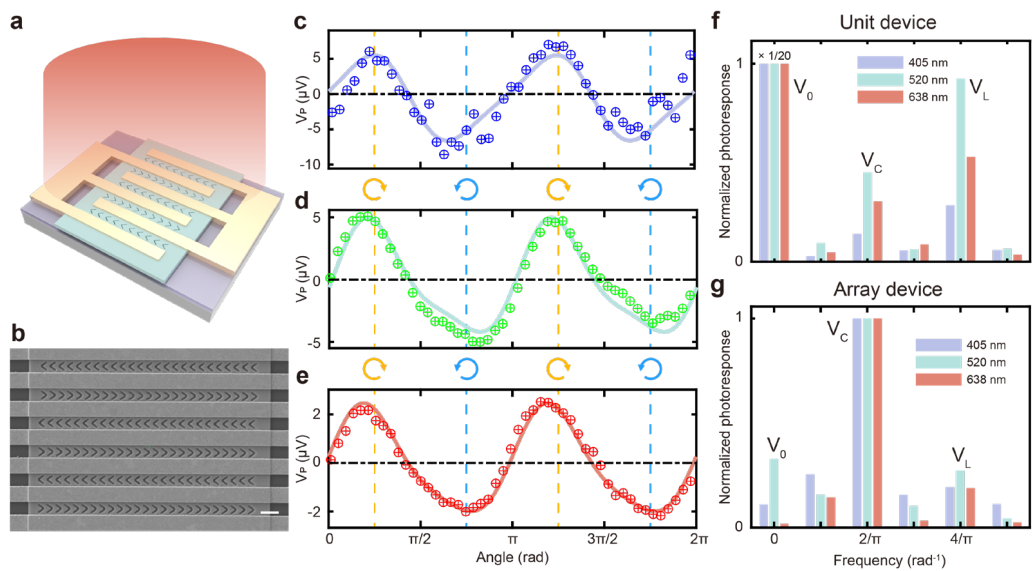

**Fig. 4 Optoelectronic characterization of optimized array devices. a.** Scheme of an array device excited by uniform large spot. **b.** SEM image of an array device. **c-e.** Variation curves of photovoltage with QWP rotation angle at wavelengths of 405 nm, 520 nm, and 638 nm. Yellow and blue vertical lines represent RCP and LCP lights. **f.** Photoresponse spectrum diagram of unit devices under uniform field excitation. Results for each wavelength are normalized to polarization-insensitive terms. **g.** Photoresponse spectrum diagram of array devices under uniform field

excitation, with results at each wavelength normalized to CPL-sensitive terms. The components at angular frequencies 0, 2/π, and 4/π represent $V_0$, $V_C$, and $V_L$ terms.

Based on the above analysis, array devices were designed and fabricated to further reduce the $V_0$ and $V_L$ terms, achieving an impressive discrimination ratio across the visible band. This innovative design reduces the necessity for precise alignment of the light spot, enhancing the device's tolerance to spot position variations. The $V_0$ component of photovoltage arises from electrode asymmetries due to variations in contact properties, fabrication imperfections, and other factors. While $V_L$ mainly originates from $V_0$ modulation via alternate absorption. Even with uniform illumination, the electrodes on either side of a unit device may respond differently, leading to a net photovoltage, as detailed in Supplementary Figure S8. This random asymmetry can be neutralized by employing multiple electrode pairs, reducing the polarization-insensitive photovoltages and the corresponding linear polarization-dependent terms. The refined device structure, along with its SEM images, is presented in Fig. 4a, b. The device incorporates interdigitated electrodes flanked by V-shaped grooves, with adjacent rows of V-grooves oriented in opposite directions. This configuration guarantees that one side of the interdigitated electrodes consistently experiences a higher near-field mode heating under CPL. The experiments employed a broad, uniform light spot to cover the entire device, coupled with the rotation of QWP to capture the polarization-dependent photovoltage curves. Results at the wavelengths of 405 nm, 520 nm, and 638 nm are shown in Fig. 4c-e. Compared to unit devices under a focused light spot in an equilibrium state, the optimized devices exhibited suppressed polarization-insensitive and linear terms, achieving a maximum DR of 107 at 405 nm wavelength. Furthermore, satisfactory DR values were also obtained at 520 nm (20) and 638 nm

(32). The CPL-sensitive photovoltage dominates the responses, and high-precision CPL distinguish is achieved. Fig. 4f, g displays the normalized spectral distribution of the photovoltage under a uniform light field for unit and array devices, respectively. The components at angular frequencies 0, 2/π, and 4/π represent $V_0$, $V_C$ and $V_L$ terms. In the unit device, the random asymmetry-induced polarization-insensitive term completely masks the CPL-sensitive photovoltage, leading to mere magnitude variations without directional distinction under opposite CPL illumination (Supplementary Figure S11). To attain a notable discrimination ratio, it is necessary to employ a focused light spot and align it to the equilibrium state position, thereby reducing the $V_0$ term and emphasizing the Vc term. Conversely, in the array device, the CPL-sensitive term dominates under uniform excitation, substantially reducing spatial alignment demands. It is believed that further increases in the number of interdigitated fingers can lead to additional suppression of the $V_0$ and $V_L$ terms in the array devices, resulting in enhanced DR. Therefore, we have successfully demonstrated a high DR, broadband CPL photodetector using all-dielectric nanostructures. Array devices, with further improved DR and reduced alignment requirements, have emerged as promising candidates for on-chip CPL detection.

The array device also exhibits exceptional noise levels and response speed. It showcases a rise time of 290 μs and a fall time of 600 μs, yielding an estimated -3dB bandwidth of 688 Hz (Supplementary Figure S12). Previous reports suggest that the photothermoelectric effect in Te can achieve much faster response speeds[38]. The parasitic capacitance within the electrodes and the electrical measurement systems are to blame for the reduced response speed, which can be significantly enhanced through design optimization. Operating without bias voltage, the device demonstrates a remarkably low noise level, reaching down to 28 nV Hz$^{-1/2}$ around its cut-off frequency, with a noise-equivalent power (NEP) of merely 0.60 μW Hz$^{-1/2}$ (Supplementary Figure S13). In addition, our CPL photodetectors demonstrate remarkable atmospheric stability, with their performance remaining virtually unchanged after a 3-month exposure.

## Discussion

To the best of our knowledge, this work pioneers the chirality-resolving near-field modes in achiral all-dielectric metasurfaces, developing a high discrimination ratio CPL photodetector in the visible band. Distinct from conventional absorption-based CPL photodetectors, our scheme characterizes light field chirality via the polarity of photoresponse rather than the photoresponse magnitude, facilitating operation without intensity calibration. Thanks to the directional difference of photovoltage under LCP and RCP excitation, the achiral structure-based device's DR overwhelms its counterparts by two orders with a broadband response capability. These advantages allow our achiral structure-supported all-dielectric CPL detector to approach the performance of traditional CPL detectors while retaining the benefits of compactness and high integrability.

This circular polarization detection scheme based on achiral dielectric structures is not limited to specific optoelectronic materials, geometric structures, photoresponse mechanisms, or operating wavelength bands, facilitating a broad range of applications. Chirality-resolved near-field modes are widely present in achiral structures. Once this near-field chirality is converted into electrical signal readout through mechanisms such as PTE, PVE, etc., a calibration-free broadband CPL detector can be realized. For example, achiral structures on silicon can also result in uneven near-field modes to asymmetrically illuminate the electrodes (Supplementary Figure S14), which could easily convert into photocurrents directed by light chirality through Schottky junctions. The advantages of simplicity, high discrimination, and broadband response make our device suitable for a wide range of applications, including highly integrated circularly polarized cameras and multi-channel optical communications involving polarization information.

In conclusion, we have explored the chirality-resolving near-field modes in dielectric achiral structures to create a high discrimination ratio, broadband CPL photodetector in the visible band. An achiral structure-based CPL detection scheme was demonstrated with photoresponses directions determined by the chirality of the light field, thus enabling high-performance CPL distinguishing free from light intensity calibration. The adaptability of this design methodology extends its applicability to a wide range of materials, structures, and mechanisms, showcasing significant promise across various technological fields.

## Methods

**Materials synthesis and characterization**

The Te nanosheets are synthesized using a hydrothermal method involving the reduction of sodium tellurite ($Na_2TeO_3$) by hydrazine hydrate ($N_2H_4 \cdot H_2O$) at 180°C for 6 hours. Initially, 0.046g of $Na_2TeO_3$ and 2.175g of polyvinylpyrrolidone (PVP) are dissolved in 16 mL of distilled water. Subsequently, the solution is transferred to a sealed Teflon-lined autoclave, to which 1.6mL of 80% hydrazine hydrate ($N_2H_4 \cdot H_2O$) and 0.84mL of 25% ammonia solution ($NH_3 \cdot H_2O$) are added. The mixture is then gradually heated to 180°C for 6 hours. After cooling naturally, the products are washed multiple times by centrifuging with distilled water at 3000 rpm, typically 3 to 5 cycles, to yield silver-gray Te nanosheets.

The as-prepared Te nanosheets were characterized by polarized Raman spectra through a commercial spectrometer (LabRAM HR Evolution, Horiba) under a 532 nm laser. The dimensions of the sample were determined using an optical microscope (Zeiss), while its thickness was measured with Atomic Force Microscopy (AFM, SmartSPM-1000, Horiba). See AFM images in Supplementary Figure S15.

**Device Fabrication**

The Te nanosheets were dropped from the solution onto a SiO2(285 nm)/Si wafer, which had been pre-patterned with markers. After the transfer, electrodes were created on the chips using standard electron-beam lithography (ELS-F125, Elionix) with 950-PMMA resist, followed by the deposition of a 5-nm titanium (Ti) layer and a 150-nm gold (Au) layer through electron beam evaporation (DE400, DE Technology), and concluding with a lift-off process. V-grooves were then etched using an SEM-FIB system (Cross Beam 540, Zeiss), which also provided the electron microscopy images of the samples. The groove depth was controlled by adjusting the ion beam dose. AFM (SmartSPM-1000, Horiba) was employed to determine the depth of the etched grooves. (Supplementary Figure S16).

**Optoelectronic Characterization**

The photocurrent mapping was conducted through the ScanPro system (MetaTest, Nanjing). Laser beams from semiconductor lasers (MetaTest, Nanjing) with wavelengths of 405 nm, 520 nm, and

638 nm were directed through a polarizer and an achromatic QWP (LBTEK), then focused onto the devices using a 50X objective lens. QWP was automatically rotated by a stepper motor. Focused spot illumination was achieved using single-mode lasers, while multi-mode lasers provided broad, uniform illumination. The devices were mounted on a stage controlled by two piezo nanopositioners (COREMORROW) for precise positioning in the x and y directions. Laser power was measured with a power meter (PM101, THORLABS).

The photocurrent was collected by a dual-channel picoammeter (6482, Keithley), which also functioned as a voltage source for electrical characterizations. Polarization-dependent photovoltage testing and noise measurement were performed with a lock-in amplifier (MFLI, Zurich Instrument). It was controlled by a square wave signal generated by a signal generator (DG822, RIGOL Technologies) and synchronized with the semiconductor laser. The rise time and the fall time of the photoresponse were characterized by an oscilloscope (4262, Pico Technology).

## Acknowledgements

This work was supported by the Guangdong Major Project of Basic and Applied Basic Research (Grant No. 2020B0301030009), the National Key Research and Development Program of China (Grant No. 2022YFA1604304), and the National Natural Science Foundation of China (Grant No. 92250305).

## Author Contributions

G.L. conceived the idea and designed the experiments. G.Z carried out the numerical simulations, performed the device fabrication, conducted the optical and electrical characterization, and wrote the manuscript. X.L. synthesized the Te nanosheets under the guidance of Y.G. Y.Q performed the near-field measurements. Y.L. and Y.C. guided the device optimization. X.M. helped with the numerical calculation. Z.F., X.M., Y.X., and Z.C. participated in the optical and electrical characterization. Y.L., G.L., D.S. and Q.G. supervised the project, discussed the progress and results, and edited the manuscript. All authors commented on the manuscript.

## Conflict of interest

The authors declare no competing interests.